\documentclass{emulateapj}
\usepackage[dvipsnames]{color} 
\newcommand{\bl}[1]{\mbox{\boldmath$ #1 $}}
\newcommand{\be}{\begin{equation}}
\newcommand{\ee}{\end{equation}}

\slugcomment{Submitted to Astrophysical Journal}
\shorttitle{T Tauri and Brown Dwarf Disk Accretion}
\shortauthors{E. I. Vorobyov and Shantanu Basu}
\begin{document}

\title{The Bimodality of Accretion In T Tauri Stars and Brown Dwarfs}

\author{E. I. Vorobyov\altaffilmark{1,}\altaffilmark{2},
Shantanu Basu\altaffilmark{3}}
\altaffiltext{1}{Institute for Computational Astrophysics, Saint Mary's University,
Halifax, B3H 3C3, Canada; vorobyov@ap.smu.ca.} 
\altaffiltext{2}{Institute of Physics, South Federal University, Stachki 194, Rostov-on-Don, 
344090, Russia.}
\altaffiltext{3}{Department of Physics and Astronomy, University of Western Ontario,
London, Ontario, N6A 3K7, Canada; basu@astro.uwo.ca.}

\begin{abstract}{}
We present numerical solutions of the collapse of prestellar cores
that lead to the formation and evolution of circumstellar disks.
The disk evolution is then followed for up to three million years.
A variety of models of different initial masses and rotation rates
allows us to study
disk accretion around brown dwarfs and low-mass T Tauri stars,
with central object mass $M_* < 0.2\,M_{\odot}$, as well
as intermediate and upper-mass T Tauri stars
($0.2\,M_{\odot} < M_* < 3.0\,M_{\odot}$). Our models include
self-gravity and allow for nonaxisymmetric motions. In addition
to the self-consistently generated gravitational torques, we
introduce an effective turbulent $\alpha$-viscosity with $\alpha = 0.01$,
which allows us particularly to model accretion in the low-mass
regime where disk self-gravity is diminishing. A range of models with 
observationally-motivated values of the initial ratio of rotational to 
gravitational energy yields a correlation between mass accretion
rate $\dot{M}$ and $M_*$ that is relatively steep, as observed.
Additionally, our modeling reveals evidence for a bimodality in
the $\dot{M}$--$M_*$ correlation, with a steeper slope at lower masses
and a shallower slope at intermediate and upper masses, as
also implied by observations.
Furthermore, we show that the neglect of disk self-gravity leads
to a much steeper $\dot{M}$--$M_*$ relation for intermediate
and upper-mass T Tauri stars.  This demonstrates that
an accurate treatment of global self-gravity is essential to
understanding observations of circumstellar disks.

\end{abstract}


\section{Introduction}

Using numerical hydrodynamic simulations, we have recently shown that 
self-gravitating disks of intermediate- and upper mass T Tauri stars 
(TTSs) settle into a self-regulated state, with low-amplitude density 
perturbations persisting for at 
least several Myr \citep{VB07}. These perturbations are sustained by 
swing amplification at 
the disk's outer edge. The associated gravitational torques can drive mass 
accretion rates that are of correct magnitude to explain the observed values
across the range of intermediate and upper mass TTSs
\citep{VB08}. These results were used to argue that the empirically
observed correlation between mass accretion rate $\dot{M}$ and central
object mass $M_*$, of the approximate form $\dot{M} \propto M_*^2$
\citep[e.g.][]{Muzerolle1,Muzerolle2,Natta04,Calvet,Mohanty},
can be explained purely on the basis of self-regulated accretion by
gravitational torques in self-gravitating
circumstellar disks. Nonaxisymmetric structure (and therefore 
nonaxisymmetric modeling) is required to achieve 
this result. 

However, it is important to note that the 
approximate $\dot{M} \propto M_*^2$ empirical relation was only noticed
after data were obtained for brown dwarfs (BDs) and combined with 
the TTS data. The larger dynamic range of masses
and accretion rates in a combined plot reveals a relatively steep overall
gradient in $\dot{M}$ versus $M_*$. Furthermore, there is
evidence that the correlation is steeper at lower masses, and shallower
at higher masses \citep{VB08}, although this is complicated by the
fact that different techniques are usually employed to determine $\dot{M}$
in the two mass regimes \citep{Muzerolle1,Mohanty}.

In this paper we model {\em both} the lower and upper mass end of the 
$\dot{M}$--$M_*$ correlation, using additional models for BD accretion
that start from the collapse of very dense and compact low-mass cores. The low-mass
disks that form around low-mass objects in
our simulations are nearly axisymmetric and sustain negligible gravitational torques.
Therefore, we explore here the 
possibility that additional processes are at play in the low-mass regime.
These processes may depend on instabilities in the low-mass disks that
generate turbulence that in turn yields an effective local viscosity.
Our model is therefore consistent with the idea that gravitational
torques are the driving agent of mass and angular momentum transport
in intermediate- to high mass disks, while a mechanism like the
magnetorotational instability (MRI) may dominate in low-mass disks. 
Other workers have arrived at this conclusion based on physical and
observational constraints \citep[e.g.][]{Hartmann,Kratter}.
We are also motivated by the apparent bimodal slope
of the $\dot{M}$--$M_*$ correlation, implying that two different
physical processes may be dominating disk evolution 
in the two different mass regimes. 
We note that our models are tracing the accretion processes
for radii greater than 5 AU, and we are implicitly assuming that
the inner disk maintains the same accretion rate (at least in a 
time-averaged sense) through unspecified processes. Therefore, even the
gravitational-torque-dominated disks may require processes such as the
MRI to accomplish the corresponding angular momentum transport in the
inner disk \citep[see][]{Armitage,Zhu}.

We model the effect of turbulent viscosity with the $\alpha$-prescription 
\citep{SS}, implemented in a nonaxisymmetric disk,
as laid out in \S \ref{model}.
It is applied to models of all masses.
Protostellar disks are formed from the self-consistent collapse of 
initially slowly-rotating prestellar cores. The evolution
of the disks are then followed for up to three million years.
As explained in our previous papers \citep{VB05,VB06}, this is made possible
through the use of the thin-disk approximation and our nonuniform
grid. Such long-term calculations, including wide parameter surveys,
remain out of reach for fully three-dimensional calculations.
We use our results to compare the numerically-derived mass accretion 
rates with those inferred observationally 
for TTSs and BDs in the late stage of star formation.


\section{Model description}
\label{model}

\subsection{Basic Equations}

Our numerical model is similar to that used recently to simulate the secular 
evolution of viscous and self-gravitating circumstellar disks \citep{VB09}. Here, we 
briefly provide the basic concepts and equations. We use the thin-disk
approximation to compute the evolution of rotating, gravitationally bound cloud 
cores.
The numerical integration is started in the prestellar phase, which is 
characterized by a collapsing {\it starless} cloud core, and continues into the 
late accretion phase, which is characterized by a protostar/disk/envelope 
system. This ensures a {\it self-consistent} formation of circumstellar disks in 
our numerical simulations. Once the disk has formed, it occupies the innermost 
regions of our numerical grid, while the envelope occupies the rest 
of the grid. 
This means that the mass infall rate onto the disk is actually 
determined by the dynamics of gas in the envelope, rather than being
input through an ad hoc source term. 
The thin-disk approximation is an excellent means to calculate the evolution
for many orbital periods and many model parameters. It is well justified 
as long as
the ratio of the disk scale height $Z$ to radius $r$ does not considerably 
exceed 0.1. As one of us has recently shown \citep[see figure 7 in][]{Vor09}, 
this condition
is fulfilled for solar mass stars having disks of several hundred AU in radius.
Our model disks rarely exceed this size, and hence we believe that the thin-disk approximation
is justified after the disk formation epoch. This approximation is also
reasonable for the prestellar phase, since protostellar cores are found to be disk-like 
\citep{Jones01,Jones02,Goodwin02,Tassis07}.

The basic equations of mass and momentum transport in the thin-disk approximation are
\begin{eqnarray}
\label{cont}
 \frac{{\partial \Sigma }}{{\partial t}} & = & - \nabla _p  \cdot \left( \Sigma \bl{v}_p 
\right), \\ 
\label{mom}
 \Sigma \frac{d \bl{v}_p }{d t}  & = &  - \nabla _p {\cal P}  + \Sigma \, \bl{g}_p + 
 (\nabla \cdot \mathbf{\Pi})_p \, ,
\end{eqnarray}
where $\Sigma$ is the mass surface density, ${\cal P}=\int^{Z}_{-Z} P dz$ is the vertically-integrated
form of the gas pressure $P$, $Z$ is the radially and azimuthally varying vertical scale height,
$\bl{v}_p=v_r \hat{\bl r}+ v_\phi \hat{\bl \phi}$ is the velocity in the
disk plane, $\bl{g}_p=g_r \hat{\bl r} +g_\phi \hat{\bl \phi}$ is the gravitational acceleration 
in the disk plane, and $\nabla_p=\hat{\bl r} \partial / \partial r + \hat{\bl \phi} r^{-1} 
\partial / \partial \phi $ is the gradient along the planar coordinates of the disk. 
The gravitational acceleration $\bl{g}_p$ includes both the gravity of a central point object 
(when formed)
and the self-gravity of a circumstellar disk and envelope. The latter component is found 
by solving the Poisson integral using the convolution theorem \citep[see][for more details]{VB06}.
The viscous stress tensor is
\begin{equation}
\mathbf{\Pi}=2 \Sigma\, \nu \left( \nabla v - {1 \over 3} (\nabla \cdot v) \mathbf{e} \right),
\end{equation}
where $\nabla v$ is a symmetrized velocity gradient tensor, $\mathbf{e}$ is the unit tensor, and
$\nu$ is the kinematic viscosity. Equation~(\ref{mom}) describes the motion of a viscous fluid in
the most general form. This equation can be reduced to the usual equation for the conservation 
of angular momentum of a radial annulus in the axisymmetric accretion disk.
The components of $(\nabla \cdot \mathbf{\Pi})_p$ in polar coordinates ($r,\phi$) 
can be found in \citet{VB09}. 

It is well known that standard collisional viscosity (molecular viscosity) 
is negligible in application to circumstellar disks.
An alternative is an effective 
turbulent viscosity induced by instabilities in the disk. A prime
candidate is the MRI, although other mechanisms of nonlinear 
hydrodynamic turbulence cannot 
be completely ruled out due to the large Reynolds numbers involved. 
We make no specific assumptions about the source of turbulence and 
parameterize the magnitude of turbulent viscosity using the usual 
$\alpha$-prescription \citep{SS}
\begin{equation}
\label{viscosity}
\nu = \alpha \, \tilde{c}_{s} \, Z,
\end{equation}
where $\tilde{c}^2_{s}=\partial {\cal P} /\partial \Sigma$ is the effective sound speed
of (generally) non-isothermal gas. The vertical scale height $Z$ is determined in
every computational cell and at every time step of integration using
an assumption of local hydrostatic equilibrium in the gravitational field of
the central star and the disk \citep[see][]{VB09}. 
It is thus important to note that our disks are not razor-thin but have a vertical extent and 
flaring which are consistent with those predicted from detailed vertical structure models of 
irradiated accretion disks around T Tauri stars by  \citet{DAlessio}.
  
In this paper we present results for a large number of simulations with differing values of physical 
parameters but a fixed value $\alpha=0.01$. This choice is motivated by our recent analysis of the secular evolution
of viscous and self-gravitating disks \citep{VB09}. We found that {\it if} circumstellar disks
around solar-mass protostars can generate and sustain turbulence then the temporally and 
spatially averaged $\alpha$ should lie in the range $10^{-3}-10^{-2}$, 
in order to be a significant process but not completely destroy the 
self-regulated structure.
Therefore, a choice of $\alpha=0.01$ samples models in which the turbulent
viscosity has a significant and even dominant role during the late stages
of disk accretion (gravitational torques are still dominant in the early
stages for all but the lowest mass disks, as discussed in \S\ \ref{discussion}).
The possible influence of different values of $\alpha$ on our results is 
also discussed in Section~\ref{discussion}.

Equations~(\ref{cont}) and (\ref{mom}) are closed with a barotropic equation
that makes a smooth transition from isothermal to adiabatic evolution at 
$\Sigma = \Sigma_{\rm cr} = 36.2$~g~cm$^{-2}$:
\begin{equation}
{\cal P}=c_s^2 \Sigma +c_s^2 \Sigma_{\rm cr} \left( \Sigma \over \Sigma_{\rm cr} \right)^{\gamma},
\label{barotropic}
\end{equation}
where $c_s=0.188$~km~s$^{-1}$ is the sound speed in the initial state,
and $\gamma=1.4$. 
Numerical simulations with $\gamma=1.67$ produce disks that are hotter
than those with $\gamma=1.4$.
However, we have recently demonstrated  that a modest increase in disk 
temperature has an insignificant effect on the mean mass accretion rates 
\citep[see figure 8 in][]{VB09}.  
Colder disks ($T\sim40$~K at 10~AU) are more gravitationally unstable and are characterized 
by spiral modes that are of larger amplitude than those of hotter disks ($T\sim100$~K at 10~AU). 
For hotter disks, however, the amplitudes of modes also 
decrease sharply with increasing order, leaving low-order modes 
($m\le 2$) dominant.
As a result, the hotter models have less mode-to-mode interaction that
can produce some cancellation in the net gravitational torque. 
Since the low-order modes are more efficient transport 
agents, the net effect is to produce comparable mass accretion rates in disks 
that differ in temperature by a factor of two.
A similar result of increasing low-order mode dominance for hotter disks
has been found by \citet{Cai}. 
We further note that $\gamma$ depends on the
disk temperature and hence it may vary radially, with the inner disk being 
characterized by $\gamma=1.4$ but the outer disk having $\gamma=1.67$. This radial variation
in $\gamma$ is expected to decrease the radial temperature gradient in the disk
(as compared to the purely $\gamma=1.4$ disk)  but is not expected
to considerably change the mass accretion rates for the reason outlined above.  

\begin{table*}
\center
\caption{Model parameters}
\label{table1}
\begin{tabular}{ccccccc}
\hline\hline
Set & $\beta$ & $\Omega_0$ & $r_0$ & $r_{\rm out}$ & $M_{\rm cl}$ & $N$  \\
\hline
 1 & $6.2\times 10^{-4}$ & $0.28-0.47$ & $2070-3450$ & $(1.2-2.0)\times 10^4$  & $1.2-2.0$ & 5    \\
 2 & $1.3\times 10^{-3}$ & $0.27-1.33$ & $1040-5180$ & $(0.6-3.0)\times 10^4$  & $0.6-3.0$ & 11   \\
 3 & $2.5\times 10^{-3}$ & $0.37-3.40$ & $570-5180$  & $(0.3-3.0)\times 10^4$  & $0.3-3.0$ & 16   \\
 4 & $3.7\times 10^{-3}$ & $0.52-6.80$ & $350-4490$  & $(0.2-2.6)\times 10^4$  & $0.2-2.6$ & 18   \\
 5 & $5.1\times 10^{-3}$ & $0.53-10.0$ & $280-5180$  & $(0.16-3.0)\times 10^4$ & $0.16-3.0$ & 17  \\
 6 & $8.0\times 10^{-3}$ & $0.67-20.0$ & $172-5180$  & $(0.1-3.0)\times 10^4$ & $0.1-3.0$  & 18 \\
 7 & $1.2\times 10^{-2}$ & $0.80-34.0$ & $120-5180$  & $(0.07-3.0)\times 10^4$ & $0.07-3.0$ & 16 \\
 8 & $1.6\times 10^{-2}$ & $2.30-50.0$ & $95-2070$   & $(0.06-1.2)\times 10^4$ & $0.05-1.2$ & 12 \\
 9 & $2.0\times 10^{-2}$ & $1.60-91.0$ & $60-3450$   & $(0.04-2.0)\times 10^4$ & $0.03-2.0$ & 14 \\
 10& $3.2\times 10^{-2}$ & $2.0-114.0$ & $60-3450$   & $(0.04-2.0)\times 10^4$ & $0.03-2.0$ & 13 \\
 \hline
\end{tabular} 
\tablecomments{All distances are in AU, angular
velocities in km~s$^{-1}$~pc$^{-1}$, and masses in $M_\odot$.
$N$ is the number of models in each model set.}
\end{table*}

\subsection{Initial Conditions and Numerical Technique}
\label{initcond}

We start our numerical simulations from starless cloud cores, which have surface densities 
$\Sigma$ and angular velocities $\Omega$ typical for a collapsing axisymmetric magnetically
supercritical core \citep{Basu}:
\begin{equation}
\Sigma={r_0 \Sigma_0 \over \sqrt{r^2+r_0^2}}\:,
\label{dens}
\end{equation}
\begin{equation}
\Omega=2\Omega_0 \left( {r_0\over r}\right)^2 \left[\sqrt{1+\left({r\over r_0}\right)^2
} -1\right],
\end{equation}
where $\Omega_0$ is the central angular velocity, 
$r_0$ is the radial scale length defined as $r_0 = k c_s^2 /(G\Sigma_0)$ and  $k= \sqrt{2}/\pi$.
These initial profiles are characterized by the important
dimensionless free parameter $\eta \equiv  \Omega_0^2r_0^2/c_s^2$
and have the property 
that the asymptotic ($r \gg r_0$) ratio of centrifugal to gravitational
acceleration has magnitude $\sqrt{2}\,\eta$ \citep[see][]{Basu}. 
The centrifugal radius of a mass shell that encloses a mass $m$ and is 
initially located at radius $r$ is estimated to be
$r_{\rm cf} = j^2/(Gm) = \sqrt{2}\, \eta r$, where $j=\Omega r^2$ is the specific angular
momentum \citep[see][]{Basu98}. We note that $\eta$ is similar in magnitude to the ratio of
rotational to gravitational energy $\beta=E_{\rm rot}/E_{\rm grav}$, where the
rotational and gravitational energies are defined
as
\begin{equation}
E_{\rm rot}= 2 \pi \int \limits_{r_{\rm in}}^{r_{\rm
out}} r a_{\rm c} \Sigma \, r \, dr, \,\,\,\,\,\
E_{\rm grav}= - 2\pi \int \limits_{r_{\rm in}}^{\rm r_{\rm out}} r
g_r \Sigma \, r \, dr.
\label{rotgraven}
\end{equation}
Here $a_{\rm c} = \Omega^2 r$ is the centrifugal acceleration, and $r_{\rm in}$ and $r_{\rm out}$
are the inner and outer cloud core radii, respectively. The former is always set to 5~AU, while the
latter varies according to the adopted cloud core size. From here onwards, we will refer to
$\eta$ and $\beta$ as synonymous quantities. The gas has a mean molecular 
mass $2.33 \, m_{\rm H}$ and cloud cores are initially isothermal with temperature $T=10$~K.

We present results from ten sets of models, each set being characterized by a distinct value 
of $\beta$. Individual models within every set are generated by varying $r_0$ and $\Omega_0$ 
in such a way that the product $r_0 \Omega_0$ is kept constant. This ensures that $\beta$
is also constant for every model in the set, because the initial sound speed 
$c_s = 0.188$ km s$^{-1}$ is equal for all models. All models have the ratio $r_{\rm out}/r_0$
set to 6.0 to generate truncated cloud cores of similar form. 
The values of $\beta$, typical intervals for $r_{\rm out}$, $r_0$, $\Omega_0$, 
and cloud core masses $M_{\rm cl}$, and number of individual models within each set are listed 
in Table~\ref{table1}. We note that our adopted values of $\beta$ lie within
the limits inferred by \citet{Caselli} for dense molecular cloud cores: $\beta=(10^{-4} - 0.07)$.


Equations~(\ref{cont}), (\ref{mom}), (\ref{barotropic}) are solved in polar 
coordinates $(r, \phi)$ on a numerical grid with
$128 \times 128$ points. We have found that an increase in the resolution to 
$256 \times 256$ grid zones has little influence on the accretion history. 
Therefore, we use the $128 \times 128$ grid in order to save a considerable 
amount of CPU time and to find solutions for a large number of model cloud cores. 
Each model takes about 400 CPU hours on the Opteron 2.5 GHz processor.
We use the method of finite differences with a time-explicit,
operator-split solution procedure. Advection is
performed using the second-order van Leer scheme.  The radial points are logarithmically spaced.
The innermost grid point is located at $r_{\rm in}=5$~AU, and the size of the 
first adjacent cell varies in the 0.17--0.36~AU range depending on cloud core size.  
We introduce a ``sink cell'' at $r<5$~AU, 
which represents the central star plus some circumstellar disk material, 
and impose a free inflow inner boundary condition. The outer boundary is reflecting.
A small amount of artificial viscosity is added to the code, 
though the associated artificial viscosity torques were shown to be negligible 
in comparison with gravitational torques \citep{VB07}. 
For reasonable values of the $\alpha$-parameter ($\ga 10^{-4}-10^{-3}$), 
the artificial viscosity torques are also considerably smaller than the viscous torques due to
$\alpha$-viscosity. 

\section {Results}
\label{results}

\subsection{Time Evolution}

\begin{figure*}
 \centering
  \includegraphics[width=13cm]{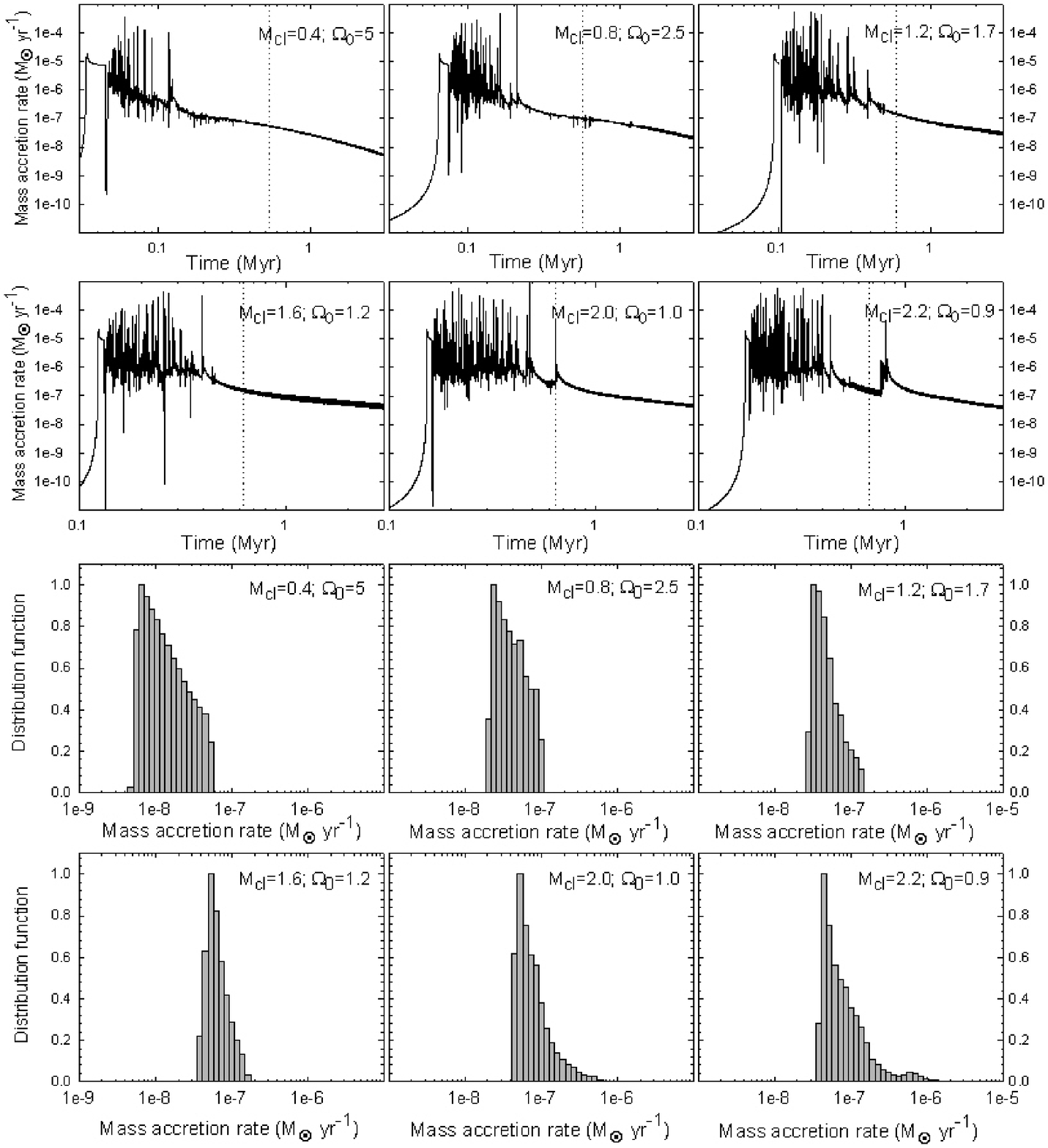}
      \caption{{\bf Two upper panels}. Mass accretion rates versus time elapsed since 
      the beginning of
      simulations. Six models from model set 6 are presented, the initial cloud core
      mass $M_{\rm cl}$ (in $M_\odot$) and central angular velocity $\Omega_0$ 
(in km s$^{-1}$ pc$^{-1}$) are indicated in every panel. Vertical 
dotted lines mark the onset of the late accretion phase at 0.5~Myr from the central object formation.
{\bf Two lower panels.} Theoretical distribution functions (DFs) of accretion rates for the same six models showing
      the normalized number of accretion ``measurements'' as a function of the mass accretion rate.
      The measurements are done every 20 yr between 0.5 Myr and 3.0 Myr from the central object formation.}
         \label{fig1}
\end{figure*}

We begin with reviewing the accretion history in objects formed from cloud cores of distinct
masses but having equal ratios of the rotational-to-gravitational energy 
$\beta$. For this purpose, we choose model set 6 with 
$\beta=8.0\times 10^{-3}$. The corresponding mass accretion rates versus 
time elapsed since the beginning of simulations
are shown in two upper rows of Figure~\ref{fig1}. The time-dependent mass 
accretion rate is calculated as $\dot{M}(t)=-2 \pi r_{\rm in} v_r \Sigma$, 
where $v_{r}$ is the inflow velocity
through the sink cell and $r_{\rm in}=5$~AU is the radius of the sink cell.
The initial cloud core mass $M_{\rm cl}$ and central angular velocity $\Omega_0$
are shown in each frame.

The early accretion history for all objects is similar---$\dot{M}$ reaches a peak value
of $\sim 2.0 \times 10^{-5}~M_\odot$~yr$^{-1}$ soon after the central object formation
and settles at a near constant value of $\sim 10^{-5}~M_\odot$~yr$^{-1}$.
Then, a transient sharp decline in $\dot{M}$ follows, manifesting the onset of
disk formation.
For a short time, centrifugal forces balance those of viscosity and gravity and the mass accretion
rate drops to a negligible value.
As the disk continues to build up its mass, the burst mode of accretion ensues. During this
phase, prolonged periods
of relatively low accretion at $10^{-7}$--$10^{-6}~M_\odot$~yr$^{-1}$ are interspersed with
short episodes of activity when $\dot{M}$ increases to 
$10^{-4}$--$10^{-3}~M_\odot$~yr$^{-1}$.
These accretion bursts are associated with disk fragmentation and formation
of dense massive clumps that are later driven onto the central
star due to the gravitational interaction with spiral arms \citep{VB06}.
Figure~\ref{fig1} demonstrates that
the intensity and duration of the burst mode increases along the line of increasing
cloud core masses. Most of the burst activity is constrained to the early several hundred
thousand years of evolution, though the most massive objects can undergo a few 
bursts even
after 0.5~Myr. In this late evolution phase, the mass accretion rates
show a moderate decline by roughly one order of magnitude during 2--3 Myr. Some short-term variability
is also present, but its magnitude is much smaller than that of the long-term
 decline.

Mass accretion rates in other model sets show a similar pattern of behavior.
We now turn to comparing our numerically derived accretion rates in the late evolution phase 
with those inferred from observations.

\subsection{Mass Accretion Rates}

In this section, we compare our model mass accretion rates and central object
masses
with those compiled by \citet[][and references therein]{Muzerolle2} for
TTSs and BDs of age 0.5--3.0~Myr based on the measurements in (mostly)
Taurus and other star formation regions. In order to facilitate the comparison,
we time-average our model mass accretion rates and central object masses
between 0.5~Myr and 3.0~Myr from the central object formation
to obtain characteristic mean values $\langle \dot{M} \rangle$ and $\langle M_{\ast} \rangle$,
respectively. For each model in Figure~\ref{fig1}, we also construct the 
distribution function (DF) of accretion rates
by calculating $\dot{M}$ every 20~yr between 0.5~Myr and 3.0~Myr from
the central object formation and
distributing the resulted values among 100 logarithmically-spaced bins in
the $10^{-9}~M_\odot$~yr$^{-1}$--$10^{-5}~M_\odot$~yr$^{-1}$ range.
The resultant DFs representing the normalized number of
accretion ``measurements'' $N_{\dot{\rm M}}$ (by analogy to observations) in a given mass
accretion bin are shown in two bottom rows of Figure~\ref{fig1}. It is seen 
that the accretion
rates in each model are localized
in a specific interval.  There is a preference for lower values of $\dot{M}$ 
in each interval,
implying the measurements of mass accretion rates may be biased towards lower
values.


\begin{figure*}
 \centering
  \includegraphics{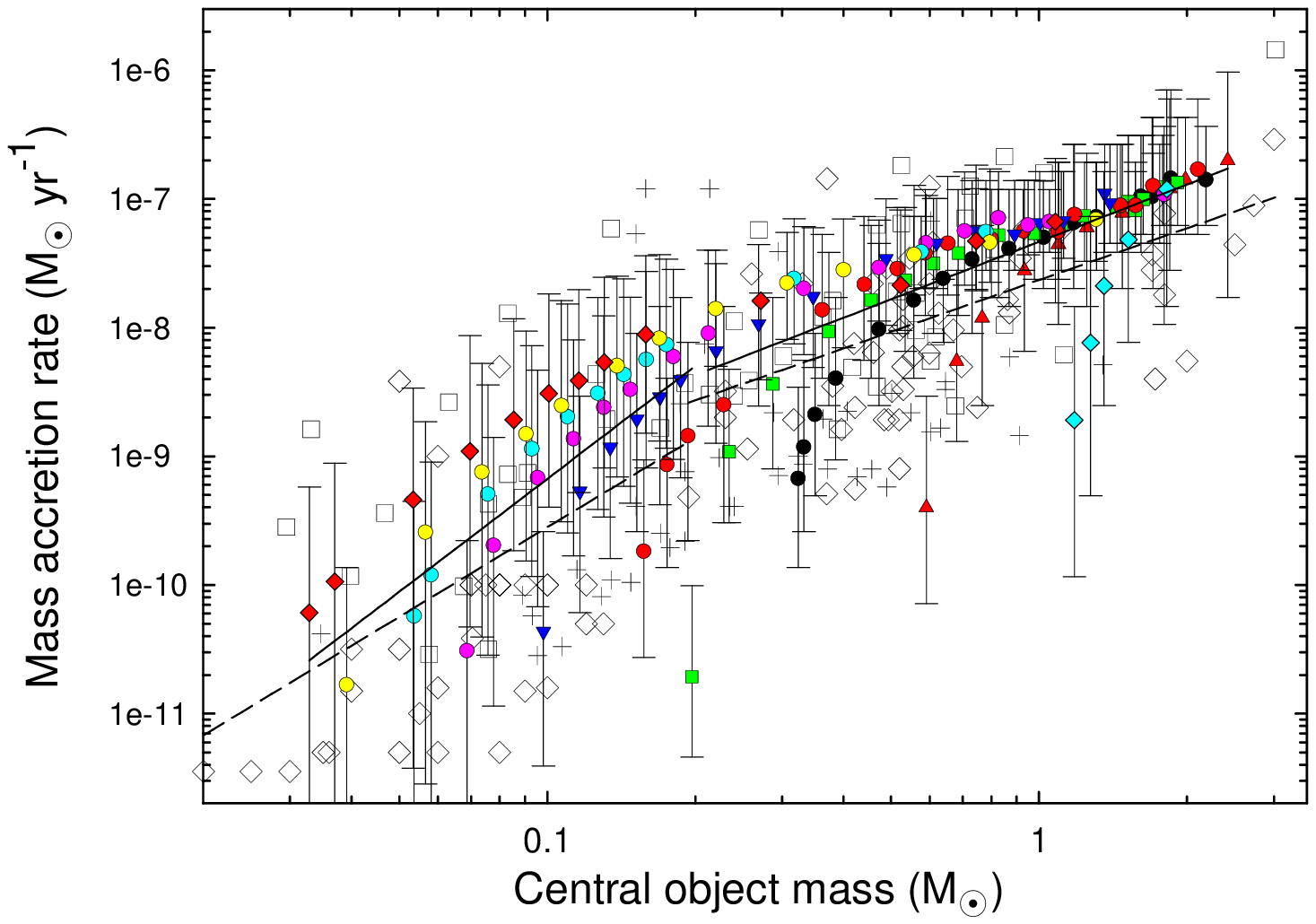}
      \caption{Mass accretion rates versus central object masses. The open
      diamonds represent measurements of TTSs and BDs from
\citet[][and references therein]{Muzerolle2}; the open squares and
plus signs represent the confirmed detections and upper limits,
respectively, compiled by \citet{Natta}.  The filled color symbols show the
time-averaged mass accretion rates $\langle \dot{M} \rangle$  versus time-averaged
central object masses $\langle M_\ast \rangle$  obtained
in our numerical modeling. Each symbol type represents a distinct set of 
models with a constant value of $\beta$, as detailed in Table~\ref{table1}.
In particular, model set 1 is plotted by cyan diamonds, model set 2---by red triangles-up,
model set 3---by black circles, model set 4---by green squares, model set 5---by red circles, 
model set 6---by blue triangles-down, model set 7---by pink circles, model set 8---by cyan circles,
model set 9---by yellow circles, and model set 10---by red diamonds. 
The bars represent typical variations in accretion rates in each model. The solid/dashed lines
are the least-squares best fits to the model/observational data, respectively. In particular,
the left lines are the fits to BDs and low-mass TTSs and right lines are the fits to
the intermediate- and upper-mass TTSs.}
         \label{fig2}
\end{figure*}

Color symbols in Figure~\ref{fig2} 
represent the time-averaged mass accretion rates $\langle \dot{M} \rangle$ 
(in $M_\odot$~yr$^{-1}$) versus time-averaged object masses $\langle M_{\ast} \rangle$ 
(in $M_\odot$). 
Each filled
symbol (of same color and shape) within a given set of models represents an individual 
object, which has formed from a cloud core of distinct mass, 
angular velocity, and size. In addition, each filled symbol is assigned vertical
bars showing typical variations in the accretion rate between 0.5~Myr and 3.0~Myr
after the central object formation. 
The typical variations are defined as mass accretion rates with $N_{\rm \dot{M
}}$ equal
to or greater than 0.05.
This procedure helps to exclude objects with accretion rates that have 
a probability
to be detected that is less than one part in twenty and are thus statistically 
insignificant.
Considering the number ($\la 20$) of detected objects
along the line of constant central object mass, we believe that our
typical variations represent conservative estimates.
The open symbols and plus signs in Figure~\ref{fig2} represent data obtained
from observations.
In particular, the open diamonds represent measurements, mostly in Taurus, that
have been compiled by \citet[][and references therein]{Muzerolle2}; the open
squares represent detections in $\rho$ Oph obtained by \citet{Natta} and plus signs
represent their upper limits to nondetections. 
We have excluded objects in the compilation of \citet{Muzerolle2}
that were later observed by \citet{Natta}.



There are several important conclusions that can be drawn by analyzing Figure~\ref{fig2}.
\begin{enumerate}
\item Our numerical models yield mass accretion rates that are of correct magnitude
to explain the observed values in TTSs and BDs.  There are only a few objects whose accretion 
rates fall beyond our predicted limits, but these objects can easily be
accounted for by increasing $\beta$ marginally above our adopted values.

\item The time-averaged mass accretion rates and stellar masses for objects from
every individual set of models form a unique track. These tracks are distinct for BDs and 
low-mass TTSs but tend to converge for upper-mass TTSs. Objects within each individual 
track have formed from cloud cores of distinct mass. In particular, 
objects in the lower-left end of the track have formed from 
cloud cores of lower mass. 


\item Our models suggest that TTSs of 0.5--3.0~Myr age may have a wider range 
of mass accretion rates than implied from observations.
In particular, we predict the existence of TTSs with accretion rates a factor of 10 lower
than has been found in the compilation of \citet[][and references therein]{Muzerolle2}. 
The lack of such objects in the observational data
is most likely explained by a difficulty to measure extremely low accretion rates in TTSs.

\item A dearth of BDs with accretion rates $\dot{M}>10^{-8}~M_\odot$~yr$^{-1}$
is likely caused by the lack of cloud cores with sufficiently 
large rotation rates. For instance, in order to reproduce the largest detected 
$\dot{M}$ in BDs ($\approx 10^{-8}~M_\odot$~yr$^{-1}$), we had to employ models 
with $\beta=0.032$. This ratio of the rotational to gravitational energy, 
according to \citet{Caselli}, is close to the upper measured limit for dense cloud cores.

\end{enumerate}

Perhaps, the most important conclusion from our modeling is that the observed scatter
in the mass accretion rates along the line of equal object masses {\it cannot}
be explained by intrinsic variability (shown in Figure~\ref{fig2} 
by vertical bars). This variability can account only for a maximum of two orders
of magnitude scatter (with the noticeable exception of a few low-mass BDs) 
and often even less, but the measured rates typically span a range 
of three orders of magnitude. Figure~\ref{fig2} clearly demonstrates that some 
object-to-object variations are necessary to explain the observed scatter.
This confirms previous claims by \citet{Natta04} and \citet{Nguyen}.

The object-to-object variations in the mass accretion rate along the line of
equal (sub)stellar masses are caused by the difference in the disk masses.
More massive disks are expected to drive higher rates of accretion \citep{VB08}.
Figure~\ref{fig2} shows that objects with greater $\langle \dot{M} \rangle$
but equal $\langle M_\ast \rangle$ belong to model sets characterized by greater
values of $\beta$. At the same time, $\beta$ controls the disk mass
because the centrifugal radius is directly proportional to $\beta$ (see Section~\ref{initcond}).
This means that models with greater $\beta$ (and greater $\langle \dot{M} \rangle$)
but equal $\langle M_\ast \rangle$ are generally expected to harbor 
more massive disks and this is confirmed by our calculations of disk masses
presented in \citet{Vor09}.

When {\it all} observational data in Figure~\ref{fig2} for TTSs and BDs are taken together, 
a least squares fit is described by a power law 
\begin{equation}
\dot{M} = 10^{-7.7} M_\ast^{1.8\pm 0.1}.
\label{relation}
\end{equation}
When we take the least-squares fit
to the confirmed detections only (excluding non-detections represented by plus signs), 
we obtain the relation 
\begin{equation}
\dot{M} = 10^{-7.5} M_\ast^{2.0\pm 0.1}.
\label{relation2}
\end{equation}
The difference between the best fits is not significant,
but it demonstrates that the exponent is sensitive to the way 
the data are handled
and may actually change when more observational data become available.

The least-squares best fit to {\it all} model data shown in Figure~\ref{fig2}
is described by the relation 
\begin{equation}
\langle \dot{M} \rangle = 10^{-7.3} \langle M_\ast \rangle^{1.8 \pm 0.1}.
\label{modelall}
\end{equation}  
It is evident that 
our numerical modeling can reproduce the observed relation reasonably well, except probably for 
the fact that we somewhat overestimate the observed rates. Possible reasons
for this small discrepancy are discussed in Section \ref{discussion}.

However, we believe that taking a best fit over the whole mass range of BDs and TTSs may be misleading.
In our previous paper \citep{VB08}, 
we reanalyzed the data obtained from observations
and argued that the exponent $n$ in the $\dot{M} \propto M_\ast^n$ relation takes different 
values for the lower-mass and upper-mass objects.
In particular, we found $n=2.3\pm0.6$ for BDs and low-mass TTSs with $M_\ast <0.25~M_\odot$, and 
$n=1.3\pm 0.3$ for intermediate- and upper mass TTSs with $0.25~M_\odot<M_\ast<3.0~M_\odot$. 
Two possible explanations were put forward for this apparent bimodality: variations in the 
observational methods and different 
mechanisms responsible for accretion. In this paper we focus on the latter alternative.

The left and right solid lines in Figure~\ref{fig2} are the least-squares fits to our model data 
for the lower- and upper-mass objects, respectively. The dashed lines 
are the corresponding best
fits to the observational data. The least-squares fits in each mass regime
are distinct and are described by the following power laws. \\
{\bf Modeling:}
\begin{equation}
\left\{ \begin{array}{ll} 
   \langle \dot{M} \rangle = 10^{-6.3} \, \langle M_\ast \rangle^{2.9\pm0.5} & \,\, 
   \mbox{if \,$\langle M_\ast \rangle < 0.2~M_\odot$, } \\ 
   \langle \dot{M} \rangle = 10^{-7.3} \, \langle M_\ast \rangle^{1.5\pm0.1} & \,\, 
   \mbox{if \,$0.2~M_\odot \le \langle M_\ast \rangle < 3.0~M_\odot$ }.  \end{array} 
   \right. 
   \label{function1}  
 \end{equation}
{\bf Observations:}
\begin{equation}
\left\{ \begin{array}{ll} 
   \dot{M}  = 10^{-7.2} \, M_\ast^{2.3\pm0.6} & \,\, 
   \mbox{if \,$ M_\ast < 0.2~M_\odot$, } \\ 
   \dot{M}  = 10^{-7.6} \, M_\ast^{1.3\pm0.3} & \,\, 
   \mbox{if \,$0.2~M_\odot \le M_\ast  < 3.0~M_\odot$ }.  \end{array} 
   \right. 
   \label{function2}  
 \end{equation}

It is evident that our numerical modeling corroborates the presence of bimodality 
in the observed $\dot{M}$--$M_\ast$ relation, though predicting a steeper dependence
of $\dot{M}$ on $M_\ast$ for BDs and low-mass TTSs. 
We note, however, that a shallower dependence in the low-mass regime 
could be obtained by expanding our modeled range of values of $\beta$.

\section{Discussion}
\label{discussion}

How can different values of $\alpha$ affect our model accretion rates shown in 
Figures~\ref{fig1} and \ref{fig2}? Let us first focus on the intermediate- and upper mass 
TTSs with 
$0.2~M_\odot \le M_\ast <3.0~M_\odot$.  T Tauri disks are unlikely to
sustain $\alpha\ga0.1$ {\it throughout the whole disk volume} during a typical disk lifetime of 
3--5 Myr. Such large values of $\alpha$ lead to a disk depletion and disappearance on time scales 
of less than 1 Myr \citep{VB09}. The results of that study excludes large values 
of $\alpha$ and leaves us with
$\alpha\le10^{-2}$ as the most probable range of values. 
The accretion rates in the $\alpha=0$ case were presented by
\citet{VB08}.
The $\langle \dot{M} \rangle\propto\langle M_\ast \rangle^n$ relation in this case was 
found to have an exponent $n=1.7\pm 0.1$, which is similar to the value found in the present
study (see Equation~\ref{function1}), but the values of $\langle \dot{M} \rangle$ were 
approximately a factor of 2 smaller than in the $\alpha=10^{-2}$ case. This means that the 
actual value of $\alpha$ will not dramatically change the 
mean mass accretion rates, as long as it is $\le 10^{-2}$.


The effect of varying $\alpha$ in the BD and low-mass TTS regime 
is more difficult to assess. Circumstellar disks
in this mass regime have a lower surface density (than disks around intermediate- and upper-mass TTSs)
 and hence are expected to be more MRI-active. On the other hand, we do not expect 
$\alpha$ to be equal to or greater than $0.1$ due to the same reasons as discussed above.
The fact that the least-squares fits to the model and
observational data in Figure~\ref{fig2} tend to converge at the low-mass end implies that 
$\alpha=0.01$ is likely to be a good choice, certainly for the BD mass regime. 
The limit of small $\alpha \la 10^{-3}$ is difficult to examine due to numerical reasons.
Normally, as in the case of intermediate- and upper-mass TTSs, disks are relatively massive and 
artificial viscosity torques are much smaller 
than gravitational ones \citep{VB07}. However, when both $\alpha$ and disk masses are small, 
artificial viscosity torques may become comparable to
both gravitational and viscous ones. In addition, the imperfections at the inner
inflow boundary could also introduce some low amplitude noise in the inner regions,
which may not be negligible in the limit of small $\alpha$ and small disk masses.
In this situation, the model mass accretion rates may be artificially overestimated. 
To avoid these complications, we decided to use 
$\alpha=10^{-2}$ in our modeling.

\begin{figure}
  \resizebox{\hsize}{!}{\includegraphics{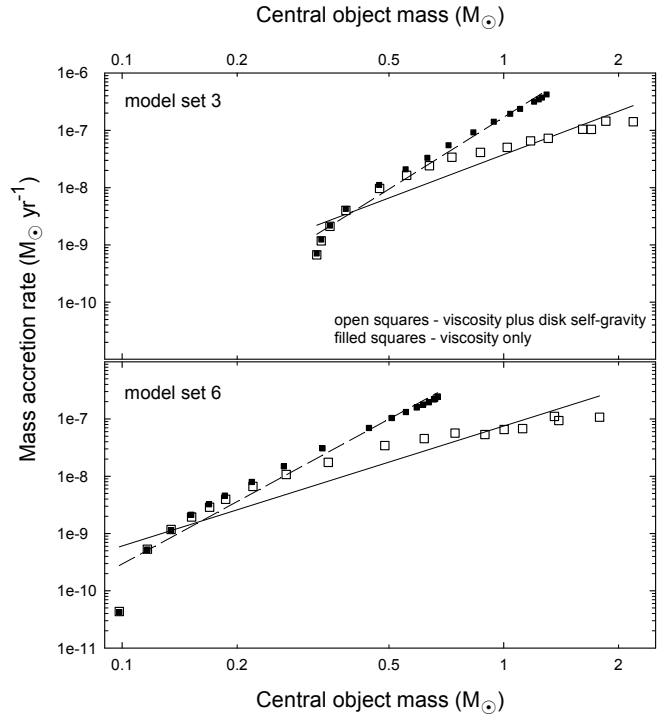}}
      \caption{Time-averaged mass accretion rates $\langle \dot{M} \rangle$ versus time-averaged
      central object masses $\langle M_\ast \rangle$ for all models from model set 3 (top)
      and model set 6 (bottom). Open squares
      show the case when both disk self-gravity and viscosity are at work, while filled squares
      correspond to the case when disk self-gravity is set to zero and viscosity remains the only
      mass transport agent in the disk. The solid and dashed lines are the least-squares best fits
      to models with disk self-gravity and without disk self-gravity, respectively.}
         \label{fig3}
\end{figure}

In the early disk evolution at $t<0.5$~Myr from the central star formation, 
gravitational-instability-induced torques 
dominate viscous ones, for 
intermediate and upper-mass TTSs \citep[see e.g.][]{VB09},
and this can influence considerably the subsequent time behavior of the system. 
We illustrate this phenomenon in Figure~\ref{fig3}, which shows time-averaged mass accretion
rates $\langle \dot{M} \rangle$ versus time-averaged central object masses 
$\langle M_\ast \rangle$ for model set~3 (top) and model set 6 (bottom). 
Every model in these sets is run twice: first time---with both
disk self-gravity and viscosity  and second time---with disk viscosity only.
In the latter case, the disk self-gravity
is artificially set to zero, thus nullifying the gravitational instability as well, 
but the gravity of the central object is kept.
The open squares in Figure~\ref{fig3} show the data obtained in models
with both disk self-gravity and viscosity (these data are identical to those shown in Figure~\ref{fig2}),
while the filled squares present the data for models with disk viscosity only. 

It is evident that gravitational instability has little effect on
the low-mass objects in each model set, but its effect becomes noticeable 
along the line of increasing object masses.
Starting from the fourth or fifth least massive object in each set, the models without disk 
self-gravity begin to yield an $\langle \dot{M} \rangle$ that is greater than those of self-gravitating
models, while at the same time underestimating the central object mass $\langle M_\ast \rangle$. 
In general, models without self-gravity tend to produce stars with masses that are too
low and mass accretion rates that are too high. This effect may seem counterintuitive since
both viscous and gravitational torques are expected to act together to increase $\langle \dot{M} \rangle$.
However, it is important to keep in mind that our time-averaged mass accretion rates 
apply to the late
evolution phase between 0.5~Myr and 3.0~Myr after the formation of the protostar. 
In contrast, during the 
early phase ($<0.5$~Myr), the time-averaged mass accretion rates
in models with self-gravity are greater than in models without self-gravity.
This is due to vigorous gravitational instability that acts in self-gravitating disks.
As a result, models with self-gravity build up the stellar mass and deplete the total gas reservoir
much faster than models without self-gravity. 
This causes the mass accretion rate in self-gravitating disks
to be lesser in the late evolution phase.

In order to better illustrate the difference in the mass accretion rates,
we take the least-squares best fits to the models with self-gravity (solid lines) and models 
without self-gravity (dashed lines). In the former case, we obtain exponents 
$n=2.5 \pm 0.2$ (top panel, model set 3) and $n=2.1 \pm 0.2$ (bottom panel, model set 6). 
When the least-squares fit is taken for both sets of models, we obtain 
$n=2.0 \pm 0.2$. This is quite similar to the exponent found for the complete 
set of models (see eq.~[\ref{modelall}]).  
In the case without self-gravity, we obtain exponents $n=4.2 \pm 0.2$ (top panel, model set 3)
and $n=3.6 \pm 0.2$ (bottom panel, model set 6),
which are considerably larger than those for models with self-gravity.  
In this example, we have not considered all models 
due to an enormous computational load, but we do not expect this tendency to change considerably
for the complete set of models.
This allows us to conclude that if not for {\it the disk self-gravity and associated
gravitational instability in the early evolution phase}, we would have had difficulty to
recover the observed $\dot{M}$--$M_\ast$ relation.

Another important feature of the non-self-gravitating models in 
Figure~\ref{fig3} is an apparent lack of bimodality in the $\dot{M}$--$M_\ast$ relation. Almost all models within an individual set,
except for a few ones with the smallest masses of the central object,
fall onto a track that is well described by a straight line in log--log space.
This strongly suggests that the {\it bimodality is the result of gravitational instability acting 
in the early phase of disk evolution.}

\section{Conclusions}
Using numerical hydrodynamic simulations of circumstellar disk formation
and evolution, we have studied the mass accretion rates in BDs and TTSs,
focusing mainly on the late evolution phase between 0.5~Myr and 3.0~Myr from the central 
object formation.
Our numerical model involves both disk self-gravity, which is accurately computed via
the solution of the Poisson integral, 
and turbulent viscosity, the latter being described by a usual 
Shakura \& Sunyaev parameterization with the $\alpha$-parameter set to $10^{-2}$.
The theoretical data are compared against those obtained from the measurements in young star-forming
clusters \citet[][and references therein]{Muzerolle2}. We find the following.

\begin{itemize}
\item Our numerical modeling of self-consistently-formed circumstellar
disks yields mass accretion rates 
of correct magnitude to explain the observed values in TTSs and BDs of 0.5--3.0~Myr age.

\item We corroborate our previous conclusion \citep{VB08} that the dependence of 
the mass accretion rates ($\dot{M}$) on the central object mass ($M_\ast$) 
in TTSs and BDs 
can be better described by a bimodal power-law function rather than that with a single exponent.
Mass accretion rates of BDs and low-mass TTSs ($M_\ast <0.2~M_\odot$) have a steeper dependence 
on $M_\ast$ than those of the intermediate- and upper-mass TTSs 
($0.2~M_\odot \le M_\ast <3.0~M_\odot$). In particular, the least-squares fits
to our model data yield exponents $n=2.9\pm0.5$ and $n=1.5\pm0.1$ for the objects with 
mass $M_\ast <0.2~M_\odot$ and $0.2~M_\odot \le M_\ast <3.0~M_\odot$, respectively.
The corresponding fits to the observational data produce exponents $n=2.3\pm0.6$ 
and $n=1.3\pm0.3$.

\item The apparent bimodality in the $\dot{M}$--$M_\ast$ relation 
is caused by vigorous gravitational instability in the early phase 
of disk evolution. The gravitational instability serves to limit disk masses in the intermediate 
and upper-mass TTSs \citep{Vor09}, thus effectively setting an upper limit on the 
mass accretion rates 
in the late evolution and flattening the $\dot{M}$--$M_\ast$ relation in this mass regime.
Models without self-gravity greatly overestimate the observed mass accretion rates, 
while at the same time underestimating the central object masses.
As a result, the non-self-gravitating models fail to account for  the observed 
$\dot{M}$--$M_\ast$ relation, predicting a much steeper dependence of $\dot{M}$
on $M_\ast$. 

%

\item The observed large scatter in mass accretion rates along the line of equal
(sub)stellar masses is caused in part by the intrinsic variability during the
evolution of individual objects,
which may span a range of one (intermediate- and  upper-mass TTSs) to two (BDs and low-mass TTSs) 
orders of magnitude. The other part is object-to-object variations due to different 
initial conditions.
Our conclusion is in agreement with that of \citet{Natta04} and \citet{Nguyen}
based on the analysis of variability in TTSs.

\item We predict the existence of TTSs with accretion rates a factor of 10
lower than those reported by \citet[][and references therein]{Muzerolle2}. 
Our modeling also indicates  that upper mass TTSs may exhibit a few FU-Ori-like outbursts 
with accretion rates greater 
than $10^{-5}~M_\odot$~yr$^{-1}$. However, the anticipated number of such peculiar objects 
is quite small and they are not expected to break the existing $\dot{M}$--$M_\ast$ 
relation.

\end{itemize}

\acknowledgements
The authors are thankful to the anonymous referee for very helpful suggestions. 
EIV gratefully acknowledges support from an ACEnet Fellowship. 
SB was supported by a grant from NSERC. We thank the Atlantic Computational 
Excellence Network (ACEnet) and  the SHARCNET consortium for access to computational facilities.

\end{document}